\documentclass[aps,prl,twocolumn,groupedaddress]{revtex4}

\usepackage{amssymb}
\usepackage{amsmath}

\makeatletter
\def\simgt{\mathrel{\lower2.5pt\vbox{\lineskip=0pt\baselineskip=0pt
           \hbox{$>$}\hbox{$\sim$}}}}
\def\simlt{\mathrel{\lower2.5pt\vbox{\lineskip=0pt\baselineskip=0pt
           \hbox{$<$}\hbox{$\sim$}}}}
\makeatother

\begin{document}

\title{Dark Matter through the Axion Portal}

\author{Yasunori Nomura}
\author{Jesse Thaler}
\affiliation{Berkeley Center for Theoretical Physics, 
  University of California, Berkeley, CA 94720}
\affiliation{Theoretical Physics Group, Lawrence Berkeley 
  National Laboratory, Berkeley, CA 94720}

\begin{abstract}
Motivated by the galactic positron excess seen by PAMELA and ATIC/PPB-BETS, 
we propose that dark matter is a TeV-scale particle that annihilates 
into a pseudoscalar ``axion.''  The positron excess and the absence 
of an anti-proton or gamma ray excess constrain the axion mass and 
branching ratios.  In the simplest realization, the axion is associated 
with a Peccei-Quinn symmetry, in which case it has a mass around 
$360~\mbox{--}~800~{\rm MeV}$ and decays into muons.  We present 
a simple and predictive supersymmetric model implementing this scenario, 
where both the Higgsino and dark matter obtain masses from the same 
source of TeV-scale spontaneous symmetry breaking.
\end{abstract}

\maketitle


{\bf Introduction.}
Evidence for dark matter (DM) is by now overwhelming~\cite{Bertone:2004pz}. 
While the precise nature and origin of DM is unknown, thermal freezeout 
of a weakly interacting massive particle (WIMP) is a successful paradigm 
that arises in many theories beyond the standard model.  If this is 
correct, then specific DM properties can be probed through direct and 
indirect detection experiments, and pieces of the dark sector might 
even be produced at the Large Hadron Collider (LHC).

Recent indirect detection results may offer important insights into 
the dark sector.  The latest PAMELA data~\cite{Adriani:2008zr} is 
strongly suggestive of a new source of galactic positrons, bolstering 
the HEAT~\cite{Beatty:2004cy} and AMS-01~\cite{Aguilar:2007yf} 
anomalies.  An intriguing interpretation of the PAMELA excess is 
DM annihilation~\cite{Bergstrom:2008gr,Cholis:2008hb,Cirelli:2008pk,%
ArkaniHamed:2008qn}.  The rate and energy spectrum of the PAMELA 
positrons are consistent~\cite{Cirelli:2008pk} with the electron/positron 
excesses seen in the balloon experiments ATIC~\cite{ATIC} and 
PPB-BETS~\cite{Torii:2008xu}, and the spectral cutoff in these 
experiments points to a DM mass in the TeV range.

There are, however, two puzzling features in the PAMELA data.  First, 
the positron excess is not accompanied by an anti-proton excess, 
which strongly constrains the hadronic annihilation modes of the 
DM~\cite{Cirelli:2008pk}.  Unless DM is as heavy as $10~{\rm TeV}$, 
the PAMELA data disfavors DM annihilation into quarks, $W$s, $Z$s, 
or Higgses.  Second, the required annihilation cross section in the 
galactic halo is orders of magnitude larger than the thermal relic 
expectation.  Therefore, any DM interpretation of the data must 
explain both the large annihilation rate and the large annihilation 
fraction into leptons.

In this letter, we propose that DM is a TeV-scale particle that 
dominantly annihilates into a pseudoscalar ``axion'' $a$.  The axion 
mass lies above the electron or muon threshold, so that $a$ dominantly 
decays into leptons with suppressed photonic and hadronic decay modes. 
In this scenario, the electron/muon decay channels would account for 
the PAMELA excess, and the photon/pion/tau decay channels would be 
constrained by gamma ray telescopes like HESS~\cite{HESS} and 
FERMI~\cite{FERMI}.

In a typical realization of the scenario, DM is a fermion.  In this 
case, the dominant annihilation channel is not to $2a$, but to $a$ and 
a real scalar $s$.  If the scalar dominantly decays as $s \rightarrow 
a a$, each DM annihilation will contain three axions, yielding 
a distinctive semi-hard galactic positron spectrum.  The existence 
of a light scalar $s$ is also crucial to enhance the DM galactic 
annihilation rate through nonperturbative effects.

This DM scenario can arise in any theory where the DM mass is 
generated from spontaneous breaking of a global $U(1)_X$ symmetry 
under which leptons have axial charges.  In two Higgs doublet models, 
it is natural to identify $U(1)_X$ with a Peccei-Quinn (PQ) symmetry 
$U(1)_{\rm PQ}$~\cite{Peccei:1977hh}, in which case our axion is 
a heavier variant of the DFSZ axion~\cite{Dine:1981rt}.  In this 
realization, the axion must decay dominantly into muons to evade 
constraints from low energy and astrophysical experiments.

The identification of $U(1)_X$ with $U(1)_{\rm PQ}$ is particularly 
well-motived in supersymmetric (SUSY) theories, since then both the 
Higgsino and DM can obtain masses from the same source of TeV-scale 
spontaneous $U(1)_{\rm PQ}$ breaking.  As we will see, the resulting 
SUSY model is extremely simple and predictive, and leads to interesting 
phenomenology at the LHC.  We thus mainly focus on this model, although 
other possibilities are also discussed.


{\bf Basic Setup.}
To understand the main features of the scenario, we first isolate the 
fields responsible for the dominant DM phenomenology and study their 
dynamics.  A complete SUSY model will be described later.

Our starting point is a global $U(1)_X$ symmetry that is broken by the 
vacuum expectation value (vev) of a complex scalar $S$:
\begin{equation}
  S = \left( f_a + \frac{s}{\sqrt{2}} \right) e^{ia/\sqrt{2}f_a},
\label{eq:S}
\end{equation}
where $a$ is the ``axion,'' $s$ is a real scalar, and $f_a$ is the axion 
decay constant.  A vector-like fermion DM $\psi/\psi^c$ obtains a mass 
from
\begin{equation}
  \mathcal{L} = -\xi S \psi \psi^c + {\rm h.c.},
\qquad
  m_{\rm DM} = \xi f_a,
\label{eq:DM}
\end{equation}
where $\psi/\psi^c$ is a standard model (SM) singlet.  The stability 
of DM can be ensured by a vector-like symmetry acting on $\psi/\psi^c$, 
which could be a remnant of $U(1)_X$~\footnote{DM may be a Majorana 
fermion $\psi$ obtaining a mass from $\mathcal{L} = -\xi S \psi^2/2 
+ {\rm h.c.}$, in which case the stability is ensured by a $Z_2$ 
symmetry.}.

In order for $a$ to decay into leptons, SM leptons must have nontrivial 
$U(1)_X$ charges.  For example, in a two Higgs doublet model, a coupling 
of the form
\begin{equation}
  \mathcal{L} = f(S) h_u h_d + {\rm h.c.},
\label{eq:S-h_int}
\end{equation}
can force $h_u h_d$, and hence the quarks and leptons, to carry 
nontrivial $U(1)_X$ charges.  The $U(1)_X$ is then a PQ symmetry, 
and we focus on this case until the end of this letter.  If the 
coupling of Eq.~(\ref{eq:S-h_int}) is sufficiently small, it does 
not drastically affect the DM phenomenology.

Unlike the ordinary axion~\cite{Weinberg:1977ma}, the mass of $a$ cannot 
come only from pion mixing, since it would then be too light to decay 
into leptons.  We therefore need some explicit breaking of $U(1)_X$. 
Also, the potential that generates the $S$ vev will also give a mass to 
$s$. Both effects can be described phenomenologically by the mass terms
\begin{equation}
  \mathcal{L} = -\frac{1}{2} m_a^2 a^2 - \frac{1}{2} m_s^2 s^2,
\label{eq:s-a_mass}
\end{equation}
and we assume the hierarchy $m_a \ll m_s \ll m_{\rm DM}$.  This condition 
is naturally satisfied in the explicit SUSY model considered later.

The scalar field $s$ decays into $a a$ through the operator
\begin{equation}
  \mathcal{L} = \frac{1}{\sqrt{2} f_a} s (\partial_\mu a)^2,
\label{eq:L-s_aa}
\end{equation}
arising from the $S$ kinetic term.  This is typically the dominant decay 
channel for $s$.


{\bf Thermal Freezeout.}
In the above setup, the DM has three major annihilation modes
\begin{equation}
  \psi \bar{\psi} \rightarrow s s,
\qquad
  \psi \bar{\psi} \rightarrow a a,
\qquad
  \psi \bar{\psi} \rightarrow s a.
\label{eq:DM-annih}
\end{equation}
The first two modes do not have an $s$-wave channel and are suppressed 
in the $v \rightarrow 0$ limit.  Annihilation at thermal freezeout 
is therefore dominated by the third mode.

In the limit $m_s, m_a \ll m_{\rm DM}$, the $v \rightarrow 0$ annihilation 
cross section is
\begin{equation}
  \langle \sigma v \rangle_{\psi\bar{\psi} \rightarrow sa} 
  = \frac{m_{\rm DM}^2}{64 \pi f_a^4} + O(v^2).
\label{eq:sigma_v-sa}
\end{equation}
A standard thermal relic abundance calculation implies
\begin{equation}
  \langle \sigma v \rangle 
  = \frac{1}{2} \langle \sigma v \rangle_{\psi\bar{\psi} \rightarrow sa} 
  \simeq 3 \times 10^{-26}\mathrm{cm}^3/\mathrm{s},
\label{eq:thermal-xsec}
\end{equation}
so once $m_{\rm DM}$ is constrained by future ATIC data, then $f_a$ is 
completely determined.  As a fiducial value, $m_{\rm DM} \sim 1~{\rm TeV}$ 
implies $f_a \sim 1~{\rm TeV}$, and so $\xi \sim 1$.


{\bf Halo Annihilation.}
The cross section of Eq.~(\ref{eq:thermal-xsec}) is too small to account 
for the observed PAMELA excess.  For $m_{\rm DM} \sim 1~{\rm TeV}$, the 
required boost factor is~\cite{Cholis:2008hb,Cirelli:2008pk}
\begin{equation}
  \langle \sigma v \rangle_{\rm PAMELA} 
  \simeq 10^3 \langle \sigma v \rangle,
\label{eq:pamela-Be}
\end{equation}
although the precise value is subject to a factor of a few 
uncertainty.  Such a large boost factor is difficult to explain 
astrophysically~\cite{Diemand:2008in}.

However, the halo annihilation rate can be enhanced by 
nonperturbative effects associated with the light state $s$, 
with $m_s \ll m_{\rm DM}$.  The relevant effects are the Sommerfeld 
enhancement~\cite{Cirelli:2008pk,ArkaniHamed:2008qn,Hisano:2004ds} 
and the formation of DM boundstates (WIMPoniums)~\cite{Pospelov:2008jd}. 
These boost the signal by
\begin{equation}
  B \simeq c \frac{\alpha_\xi m_{\rm DM}}{m_s},
\qquad
  \alpha_\xi = \frac{\xi^2}{4\pi},
\label{eq:NE-boost}
\end{equation}
where $c$ is a coefficient which can be as large as 
$(m_s/m_{\rm DM} v_{\rm halo})^2$ if $m_s$ takes values 
that allow (near) zero-energy boundstates.  Here, $v_{\rm halo} 
\sim 10^{-3}$.  As we will see, our explicit model has $m_s \approx 
O(1~\mbox{--}~10~{\rm GeV})$.  Combined with a moderate astrophysical 
boost factor, Eq.~(\ref{eq:NE-boost}) can then account for 
Eq.~(\ref{eq:pamela-Be}).  The condition for the effects being 
operative in the halo, but not at freezeout, is $v_{\rm halo} \simlt 
\alpha_\xi \simlt v_{\rm freezeout}$, where $v_{\rm freezeout} \sim 
0.4$~\footnote{Alternatively, we can make $\psi/\psi^c$ charged under 
a new $U(1)_{\rm NE}$ gauge symmetry.  As long as the $U(1)_{\rm NE}$ 
gauge coupling is small, $g_{\rm NE} \simlt \xi$, thermal freezeout 
is unaffected and the halo boost is $B \sim \alpha_{\rm NE}/v_{\rm halo}$. 
However, $U(1)_{\rm NE}$ must be spontaneously broken to avoid constraints 
from big-bang nucleosynthesis, since axion decays would otherwise 
generate a large energy density of $U(1)_{\rm NE}$ gauge bosons. 
Nonperturbative enhancements through massless gauge bosons are also 
ruled out by astrophysical observations~\cite{Kamionkowski:2008gj}. 
The model presented in the text avoids both bounds automatically, 
since the scalar $s$ has a mass $\simgt O(1~{\rm GeV})$.}.

The DM annihilation (or para-WIMPonium decay) product  is $s a$. 
After the decay $s \rightarrow a a$, this yields three axions per DM 
annihilation.  There is also an annihilation channel into $t\bar{t}$ 
through $s$-channel $a$ exchange, but its branching fraction is only 
of $O(1\%)$.  This level of hadronic activity is consistent with the 
PAMELA data.

Since DM does not annihilate directly into leptons, the positron 
injection spectrum is different from~\cite{Cirelli:2008pk} and closer 
to~\cite{Cholis:2008vb}.  If a DM annihilation product comes from a 
(scalar) cascade decay with large mass hierarchies involving $n$ steps, 
then its energy spectrum is proportional to $\{\ln(m_{\rm DM}/E)\}^{n-1}$, 
where $n=0$ for direct annihilation.  The lepton spectrum per DM 
annihilation is then
\begin{equation}
  \frac{dN_{\ell}}{dE} 
  = \frac{2}{m_{\rm DM}} \left(2 \log\frac{m_{\rm DM}}{E} + 1 \right)
\quad
  (E < m_{\rm DM}).
\label{eq:l-spectrum}
\end{equation}
This semi-hard lepton spectrum is an unambiguous prediction of 
our setup.  Since our axion will primarily decay into muons, 
Eq.~(\ref{eq:l-spectrum}) must be convoluted with muon decays 
to give the final positron spectrum.


{\bf Axion Decays.}
To account for the observed positron excess, the axion must dominantly 
decay into leptons.  As we will see, there are strong bounds on the 
photon flux from DM annihilations and this constrains $a \rightarrow 
\gamma \gamma$ to be less than $\approx 1\%$.  Since $\pi^0 \rightarrow 
\gamma \gamma$, the decay into neutral pions must also be suppressed 
to the $5\%$ level.  This disfavors the possibility $a \rightarrow 
\tau^+\tau^-$, since tau decay leads to an $O(1)$ fraction of $\pi^0$s.

Compared to the QCD axion, we have an extra $m_a^2$ parameter which 
affects axion-pion mixing.  In terms of the mixing angle for the QCD 
axion $\bar{\theta}_{a\pi^0} \sim f_\pi / f_a$, the new mixing angle 
is $\theta_{a\pi^0} = \bar{\theta}_{a\pi^0} m_\pi^2/(m_\pi^2 - m_a^2)$. 
Close to the $\pi^0$ threshold there is resonant enhancement, but for 
$m_a^2 \gg m_\pi^2$, there is a $m_\pi^2/ m_a^2$ suppression.  A similar 
enhancement also arises for $m_a^2 \simeq m_\eta^2$.

For a generic axion, the partial widths to leptons and photons are
\begin{eqnarray}
  \Gamma(a \rightarrow \ell^+ \ell^-) 
  &=& c_\ell^2 \frac{m_a}{16\pi} \frac{m_\ell^2}{f_a^2},
\label{eq:a-ll}\\
  \Gamma(a \rightarrow \gamma \gamma) 
  &=& c_\gamma^2 \frac{\alpha_{\rm EM}^2}{128 \pi^3} \frac{m_a^3}{f_a^2}.
\label{eq:a-2gamma}
\end{eqnarray}
In our case, $c_\ell = \sin^2\!\beta$, where $\tan\beta \equiv \langle 
h_u \rangle/\langle h_d \rangle$, and $c_\gamma$ depends on $m_a$ through 
$\theta_{a\pi^0}$ but is typically $O(1)$.   Except when $m_a$ is very 
far from a lepton mass threshold, the photon branching fraction is less 
than $O(1\%)$.

The partial width to pions is more complicated.  Direct decays $a 
\rightarrow \pi\pi$ are suppressed by CP invariance, and radiative decays 
$a \rightarrow \pi\pi \gamma$ are suppressed by $\alpha_{\rm EM}/4\pi$. 
The first dangerous channel is $a \rightarrow \pi\pi\pi$ which arises 
from axion-pion mixing.  We estimate the partial width as
\begin{equation}
  \Gamma(a \rightarrow \pi\pi\pi) 
  \sim \frac{1}{128 \pi^3} \frac{m_a^5}{f_\pi^4} 
    \left(\frac{f_\pi}{f_a} \frac{m_\pi^2}{m_a^2}\right)^2,
\label{eq:a-3pi}
\end{equation}
where the combination in parentheses is the approximate axion-pion mixing 
angle when $m_a^2 \gg m_\pi^2$.  Parametrically, the $a \rightarrow 
\pi\pi\pi$ mode is two orders of magnitude suppressed compared to the 
$a \rightarrow \mu^+\mu^-$ mode.  As $m_a$ approaches the $\rho\pi$ 
threshold, $a \rightarrow \pi\pi\pi$ is enhanced by $m_\rho^2/\Gamma_\rho^2$. 
Also, $a \rightarrow \eta \pi \pi$ decay becomes important around the 
same mass scale, so we estimate the total axion to $\pi^0$ branching ratio 
to be safe for $m_a \simlt 800~{\rm MeV}$.


{\bf Axion Constraints.}
The bounds on heavy axions are different from ordinary axions. 
For the range $2m_e < m_a < 2m_\mu$, a beam-dump experiment at 
CERN~\cite{Bergsma:1985qz} looked for the decay $a \rightarrow e^+e^-$, 
and definitively rules out axion decay constants up to $f_a \sim 
10~{\rm TeV}$.

In the region $2m_\mu < m_a < m_K - m_\pi$, our axion decays into 
$\mu^+\mu^-$ with $c\tau_a \simeq O(1~\mbox{--}~10~{\rm \mu m})$, 
and measurements of rare kaon decays $K \rightarrow \pi \mu^+ \mu^-$ 
constrain the branching ratio $K \rightarrow \pi a$.  The estimated 
branching ratio is ${\rm Br}\left(K^+ \rightarrow \pi^+ a \right) 
\simgt 3 \times 10^{-6} (1~{\rm TeV}/f_a)^2$~\cite{Antoniadis:1981zw}, 
and the measured rate ${\rm Br}\left(K^+ \rightarrow \pi^+ \mu^+ 
\mu^- \right) \simeq 1 \times 10^{-7}$~\cite{Park:2001cv} is consistent 
with SM expectations.  Therefore, this region seems to be excluded for 
$f_a \sim 1~{\rm TeV}$, especially considering that the dimuon invariant 
mass distribution would be peaked at $m_a$ for the axion decay.

For $m_K - m_\pi < m_a \simlt 800~{\rm MeV}$, there are interesting 
implications for rare $\Upsilon$ decays.  The predicted rate is 
${\rm Br}(\Upsilon \rightarrow \gamma a) \simeq 3 \times 10^{-6} 
\sin^4\!\beta\, (1~{\rm TeV}/f_a)^2$, while the experimental bound 
is ${\rm Br}(\Upsilon \rightarrow \gamma a) \simlt \mbox{few} \times 
10^{-6}$ for prompt $a \rightarrow \mu^+\mu^-$ decays~\cite{Love:2008hs}. 
The region of interest $f_a \sim 1~{\rm TeV}$ will be tested in future 
$B$-factory analyses.

In summary, the allowed region for our axion is
\begin{equation}
  m_K - m_\pi < m_a \simlt 800~{\rm MeV},
\label{eq:a-mass}
\end{equation}
and the dominant decay channel is $a \rightarrow \mu^+ \mu^-$.  For 
axions as heavy as $m_K - m_\pi$, astrophysical bounds are irrelevant.


{\bf Gamma Ray Bound.}
As already mentioned, the axion typically has a non-zero branching 
fraction into photons (or $\pi^0$s), and there are important bounds 
from gamma ray experiments.  Since the photon spectrum is semi-hard, 
the strongest bounds come from atmospheric Cherenkov telescopes.  The 
expected photon spectrum also overlaps with the energy range of FERMI.

A model-independent constraint comes from a HESS study of the Sagittarius 
dwarf galaxy~\cite{Aharonian:2007km}.  They put an upper bound on the 
integrated gamma ray flux for $E_\gamma > 250~{\rm GeV}$ of $\Phi_\gamma 
< 3.6 \times 10^{-12}~{\rm cm}^{-2}\,{\rm s}^{-1}$.  From a given photon 
spectrum, this can be translated into a bound on the DM annihilation 
cross section
\begin{equation}
  \langle \sigma v \rangle 
  < \frac{4\pi \Phi_\gamma m_{\rm DM}^2}{\bar{J} \Delta\Omega} 
    \left(\int_{250~{\rm GeV}}^{m_{\rm DM}}\! 
    \frac{d N_\gamma}{d E}\, d E \right)^{-1},
\label{eq:HESS-bound}
\end{equation}
where $\bar{J} \simeq 2.2 \times 10^{24}~{\rm GeV}^2\,{\rm cm}^{-5}$ 
is the Sagittarius line-of-sight-integrated squared DM density assuming 
an NFW profile, and $\Delta \Omega = 2 \times 10^{-5}$ is the HESS 
solid angle integration region.

Since the annihilation cross section is known from Eq.~(\ref{eq:pamela-Be}), 
we can translate Eq.~(\ref{eq:HESS-bound}) into a bound on the branching 
fraction to photons.  For direct $a \rightarrow \gamma \gamma$ decays, 
the energy spectrum is proportional to Eq.~(\ref{eq:l-spectrum}), 
$dN_{\gamma}/dE \simeq {\rm Br}(a \rightarrow \gamma \gamma)\, 
dN_{\ell}/dE$, and using the fiducial $m_{\rm DM} = 1~{\rm TeV}$, 
we obtain the bound
\begin{equation}
  {\rm Br}(a \rightarrow \gamma \gamma) \simlt 1\%
\quad
  (m_{\rm DM} = 1~{\rm TeV}).
\label{eq:a-photon-bound}
\end{equation}
The bound on axion decays into pions can be derived similarly.  Assuming 
$a \rightarrow \pi^0 \pi^+ \pi^-$, we obtain
\begin{equation}
  {\rm Br}(a \rightarrow \pi \pi \pi) \simlt 5\%
\quad
  (m_{\rm DM} = 1~{\rm TeV}).
\label{eq:a-pi-bound}
\end{equation}
The decay $a \rightarrow \tau^+ \tau^-$ leads to a bound on $\langle 
\sigma v \rangle$ an order of magnitude stronger than $\langle \sigma v 
\rangle_{\rm PAMELA}$~\footnote{The bounds from photon flux may be 
weakened by assuming a different DM density or velocity profile, or 
due to the uncertainty in Eq.~(\ref{eq:pamela-Be}).  In this case, 
the region $2 m_\tau < m_a < 2 m_b$ may open up, since hadronic 
axion decays are suppressed by $3 m_c^2/m_\tau^2 \tan^4\!\beta$. 
The constraint from rare $\Upsilon$ decays is satisfied, since 
the experimental upper bound on ${\rm Br}(\Upsilon \rightarrow 
\gamma a)$ with $a \rightarrow \tau^+\tau^-$ is $\mbox{few} 
\times 10^{-5}$~\cite{Love:2008hs}.  The region $800~{\rm MeV} 
\simlt m_a \simlt 2~{\rm GeV}$ may also be allowed.}.


{\bf Supersymmetric Model.}
In a SUSY context, it is natural to assume that the vector-like DM mass 
is related to the vector-like Higgsino mass $\mu_H$.  In fact, the simple 
superpotential
\begin{equation}
  W = \xi S \Psi \Psi^c + \lambda S H_u H_d,
\label{eq:W}
\end{equation}
together with the soft SUSY breaking terms
\begin{equation}
  \mathcal{L}_{\rm soft} 
  = - \xi A_\xi S \Psi \Psi^c - \lambda A_\lambda S H_u H_d 
    - m_S^2 S^\dagger S + \cdots,
\label{eq:L_soft}
\end{equation}
have all the required ingredients, except for the origin of the axion 
mass, which we leave unspecified (can simply be a small $\kappa S^3$ 
term in the superpotential).  Without the $\xi$ terms, this model is 
the PQ-symmetric limit of the NMSSM, and is sometimes referred to as 
PQ-SUSY \cite{Ciafaloni:1997gy,Miller:2003ay}.  The vevs for $S$, 
$H_u$ and $H_d$ can be generated in a stable vacuum, giving 
$m_{\rm DM}/\mu_H = \xi/\lambda$.

\begin{table*}
\begin{center}
\begin{tabular}{|c|c|c|c||c|c|c|c|c||c|c|c||c|c|c||c|c||c|}
\hline
 $m_{\rm DM}$ & $\lambda$ & $\tan\beta$ & $m_S^2$ & 
   $f_a$ & $\mu_H$ & $A_\lambda$ & $m_{H_u}^2$ & $m_{H_d}^2$ & 
   $m_s$ & $\tau_s$ & ${\rm Br}(s \rightarrow f\bar{f})$ & 
   $m_{\tilde{s}}$ & $m_{3/2}$ & $\tau_{\tilde{s}}$ & 
   $m_a$ & $\tau_a$ & 
   $\sigma_{\rm SI}~[{\rm cm}^2]$ \\
\hline \hline
 $1000$ & $0.25$ & $2.0$ & $-6.8^2$ & 
   $1100$ & $270$ & $650$ & $110^2$ & $530^2$ & 
   $34$ & $4 \cdot 10^{-21}$ & $f=b: 3\%$ & 
   $5.5$ & $10~{\rm eV}$ & $2 \cdot 10^{-5}$ & 
   $0.7$ & $8 \cdot 10^{-15}$ & 
   $3 \cdot 10^{-43}$ \\
\hline
 $1200$ & $0.10$ & $4.0$ & $-6.3^2$ & 
   $1200$ & $120$ & $430$ & $-69^2$ & $440^2$ & 
   $5.6$ & $1 \cdot 10^{-18}$ & $f=\tau: 5\%$ & 
   $1.2$ & $5~{\rm eV}$ & $0.02$ & 
   $0.4$ & $1 \cdot 10^{-14}$ & 
   $4 \cdot 10^{-43}$ \\
\hline
\end{tabular}
\end{center}
\caption{Two sample spectra in the SUSY model.  $m_{\rm DM}$, $\lambda$, 
 $\tan\beta$, $m_S^2$, $m_{3/2}$, and $m_a$ are inputs, and the rest are 
 outputs.  All the masses are in GeV (except where indicated), and the 
 lifetimes are in seconds.  $\sigma_{\rm SI}$ is a spin-independent 
 DM-nucleon cross section.  $m_{\tilde{s}}$ is calculated assuming 
 decoupling gauginos.}
\label{table:sample}
\end{table*}
For $\lambda \ll 1$ and $|m_S^2| \ll \lambda^2 v_{\rm EW}^2$, where 
$v_{\rm EW} \simeq 174~{\rm GeV}$, the dominant phenomenology is 
determined essentially by five parameters
\begin{equation}
  \{ m_{\rm DM}, \lambda, \tan\beta, m_S^2, m_a \},
\label{eq:parameters}
\end{equation}
with $m_S^2$ and $m_a$ affecting only scalar mixing and axion decay, 
respectively.  All other parameters are either determined by thermal 
relic calculations, electroweak symmetry breaking, or are secondary 
to the phenomenology relevant here.  The model is thus extremely 
predictive in this region, and we present sample spectra in 
Table~\ref{table:sample}.

The present SUSY model introduces important additions to the minimal 
structure described before.  First, the mass of $s$ is no longer a free 
parameter and is fixed by $m_s \simeq \lambda\, v_{\rm EW} \sin(2\beta)$. 
Second, we have an additional light state $\tilde{s}$, the fermion 
component of $S$, whose mass is $m_{\tilde{s}} \simeq O(\lambda^2 
v_{\rm EW}^2/m_{\rm SUSY})$, where $m_{\rm SUSY}$ is $\mu_H$ or 
a gaugino mass.

The existence of $S$ states lighter than the electroweak scale is 
consistent with the experimental data.  These states mix with $H_{u,d}$ 
states with mixing angles of $O(v_{\rm EW}/f_a)$, and constraints 
from LEP are satisfied for $f_a \simgt 1~{\rm TeV}$.  Considering 
$\mu_H = \lambda f_a$ ($\simeq A_\lambda \sin(2\beta)/2$ from potential 
minimization), $f_a \sim 1~{\rm TeV}$ implies $\lambda \approx O(0.1)$, 
which satisfies the bound on charginos.

Small values of $\lambda$ allow $m_s \approx O(1~\mbox{--}~10~{\rm GeV})$, 
as needed for the halo annihilation enhancement.  Small $\lambda$ also 
ensures that DM annihilates mainly into $S$ states and not $H_{u,d}$ 
states, which would give more hadronic activity than is allowed by PAMELA. 
To suppress additional hadronic/photonic activity from $s$ decays, the 
branching fraction of $s$ into quarks and taus can be made smaller 
than $O(10\%)$.  The $s \rightarrow \tilde{s}\tilde{s}$ mode is 
subdominant, and annihilations of DM into $\tilde{s}\tilde{s}$ 
are velocity suppressed.

Since $s$ is light, it mediates a large DM-nucleon cross section, leading 
to tension with the direct detection bound~\cite{Ahmed:2008eu}.  There 
are two ways this bound can be satisfied.  One is to take $m_s$ to be 
a few tens of GeV.  In this case the halo annihilation enhancement occurs 
through (near) zero-energy boundstates.  The other is to suppress the 
$s$-$h_d$ mixing by taking appropriate values of $|m_S^2| \approx 0.1 
\lambda^2 v_{\rm EW}^2$.  In this case the $s$ coupling to the nucleon 
can be accidentally small, with a mild tuning of $O(10\%)$.  The two cases 
described here correspond to the two points in Table~\ref{table:sample}. 
The Sommerfeld enhancement factors for these points are $\simgt 
100$~\cite{ArkaniHamed:2008qn}.

For the consistency of the DM story, $\tilde{s}$ must not be stable. 
In low-scale SUSY breaking, it is natural to assume that $\tilde{s}$ 
decays into the gravitino $\tilde{G}$.  The mass of $\tilde{s}$ 
is typically above $m_a$, in which case the lifetime is given by 
$\tau_{\tilde{s} \rightarrow a \tilde{G}} \simeq 96\pi m_{3/2}^2 
M_{\rm Pl}^2/m_{\tilde{s}}^5$.  For a gravitino mass $m_{3/2} \simlt 
O(10~\mbox{--}~100~{\rm eV})$, this is sufficiently short that 
$\tilde{s}$ never dominates the universe.  Also, gravitinos this 
light do not cause a cosmological problem~\cite{Viel:2005qj}.

The light $s$, $a$, and $\tilde{s}$ states have interesting implications 
for LHC phenomenology.  For example, the Higgs can decay as $h \rightarrow 
aa \rightarrow 4\mu$.  Strongly produced SUSY particles will typically 
cascade decay into the light Higgsino, which will subsequently decay 
into $\tilde{s}$, sometimes by emitting an $a$ or $s$.  This leads 
to pairs and quartets of collimated muons with small invariant mass.

Finally, we note that in general, DM can either be the fermion or scalar 
component of $\Psi/\Psi^c$, depending on which is lighter.  Scalar DM 
works similarly to fermion DM, except the dominant annihilation modes 
are now $ss$ and $aa$.  Since the scalar/fermion mass splitting controls 
radiative corrections to $m_S^2$, the $\Psi/\Psi^c$ states could be 
nearly degenerate, leading to coannihilation at freezeout.


{\bf Outlook.}
We have presented a DM scenario that naturally explains the PAMELA and 
ATIC/PPB-BETS data.  DM is a TeV-scale particle annihilating into an 
axion $a$, and the halo annihilation rate is enhanced through the scalar 
$s$.  In the simplest realization, $a$ is associated with a PQ symmetry, 
and we have constructed a corresponding SUSY model where the Higgsino 
and DM masses have a common origin from $U(1)_{\rm PQ}$ breaking.

There are other implementations of our scenario.  For example, in 
models of low-scale dynamical SUSY breaking, the sector breaking SUSY 
typically leads to an $R$ axion with the decay constant $f_a \sim 
\Lambda/4\pi$, where $\Lambda \approx O(10~\mbox{--}~100~{\rm TeV})$ 
is the dynamical scale.  This axion can serve as our $a$ if the Higgses 
and DM obtain $U(1)_R$ breaking masses.  The mass of $a$ is $m_a^2 
\sim \Lambda^3/M_{\rm Pl}$~\cite{Bagger:1994hh}, and the scales 
could work with $O(1)$ (or loop) factors.

One could also consider a purely leptonic axion, for example, by 
introducing a separate $U(1)_X$ and Higgs fields for the lepton sector. 
In this case $a$ does not have hadronic couplings, eliminating the 
tension with direct detection experiments and opening the possibility 
for $a \rightarrow e^+e^-$ decays.

{\bf Acknowledgments.}
We thank Andrew Cohen, Aaron Pierce, and Martin Schmaltz for useful 
conversations.  This work was supported in part by the U.S. DOE under 
Contract DE-AC02-05CH11231, the NSF under grants PHY-0457315 and 
PHY-0555661, DOE OJI, and the Alfred P. Sloan Foundation.  J.T. is 
supported by the Miller Institute for Basic Research in Science.

\end{document}